# Rashba **Spin–Orbit Driven Topological Phase Transitions in Heterogeneous** Armchair **Honeycomb Nanoribbons**

Hao-Ru Wu, Jhih-Shih You*, Yiing-Rei Chen†, and Hong-Yi Chen‡
*Department of Physics, National Taiwan Normal University, Taipei, 116, Taiwan*

We investigate the emergence of nontrivial topological phases in heterogeneous armchair honeycomb nanoribbons arising from the interplay between structural geometry and Rashba spin–orbit coupling (RSOC). The system consists of a central RSOC-active region sandwiched between two pristine segments, forming interfaces between topologically distinct phases. As the RSOC strength increases, interface states emerge and become localized at the junctions, exhibiting robustness against edge perturbations. For finite ribbon widths, the RSOC induces a closing and subsequent reopening of the bulk energy gap, signaling a topological phase transition without altering the underlying lattice geometry. These findings reveal a route to engineering tunable topological states through the cooperative effects of interfacial structure and spin–orbit interactions.

Symmetry and topology jointly govern the emergence of protected boundary states through the bulk–edge correspondence [1-3], establishing a fundamental connection between electronic structure and wave-function geometry in quantum systems. Subsequently, a general topological invariant has been proposed to extend bulk–edge correspondence to asymmetric systems [4-6]. In low-dimensional lattices, topological phases are conventionally realized through structural modifications such as lattice deformation [7], width variation [8, 9], edge reconstruction [10-15], or hopping geometry [16] to induce nontrivial band topology. While these approaches have enabled symmetry-protected edge modes in graphene nanoribbons and related systems, they typically require permanent geometric alternations to lattice structure, thereby limiting *in situ* tunability of topological properties. Therefore, we aim to explore the mechanism of tuning topological phases without relying on geometric modification.

Spin–orbit coupling provides a natural candidate mechanism to generate and control topological phases without modifying lattice geometry. In particular, Rashba spin–orbit coupling (RSOC), originating from inversion-symmetry breaking induced by substrates [17] or external electric fields [18], provides a highly tunable interaction widely exploited in spin transport and spintronic devices [19-22]. Notably, RSOC is commonly realized to destroy topological insulating phases in 2D honeycomb lattices [23, 24] and induce nontrivial topological phases in certain quasi-1D systems [25–27]. Nevertheless, the capability of RSOC to systematically drive topology in geometrically identical structures remains unexplored.

Here we show that the interplay between structural geometry and RSOC enables tunable topological phases without modifying lattice topology. We consider a heterogeneous armchair honeycomb nanoribbon in which a RSOC-active region (R) is embedded between pristine nanoribbons (P), forming a P–R–P junction. By continuously tuning the Rashba coupling strength, the system undergoes gap closing and reopening associated with distinct topological invariants, leading to symmetry-protected interface states localized at junction boundaries. These edge modes remain robust against edge perturbations, demonstrating that RSOC acts as an effective control parameter for inducing topological phase transitions in fixed-geometry systems.

Our mechanism can be extended to photonic crystals [28–30] and other synthetic lattices [31–33], where honeycomb lattice structures can be effectively engineered and Rashba-type spin–orbit interactions, together with nanoribbon geometries, have recently been realized. Our results therefore identify spin–orbit interaction as a general mechanism for generating dynamically tunable topological states across a broad class of wave-based platforms.

* jhihshihyou@ntnu.edu.tw
† yrchen@phy.ntnu.edu.tw
‡ hongyi@ntnu.edu.tw

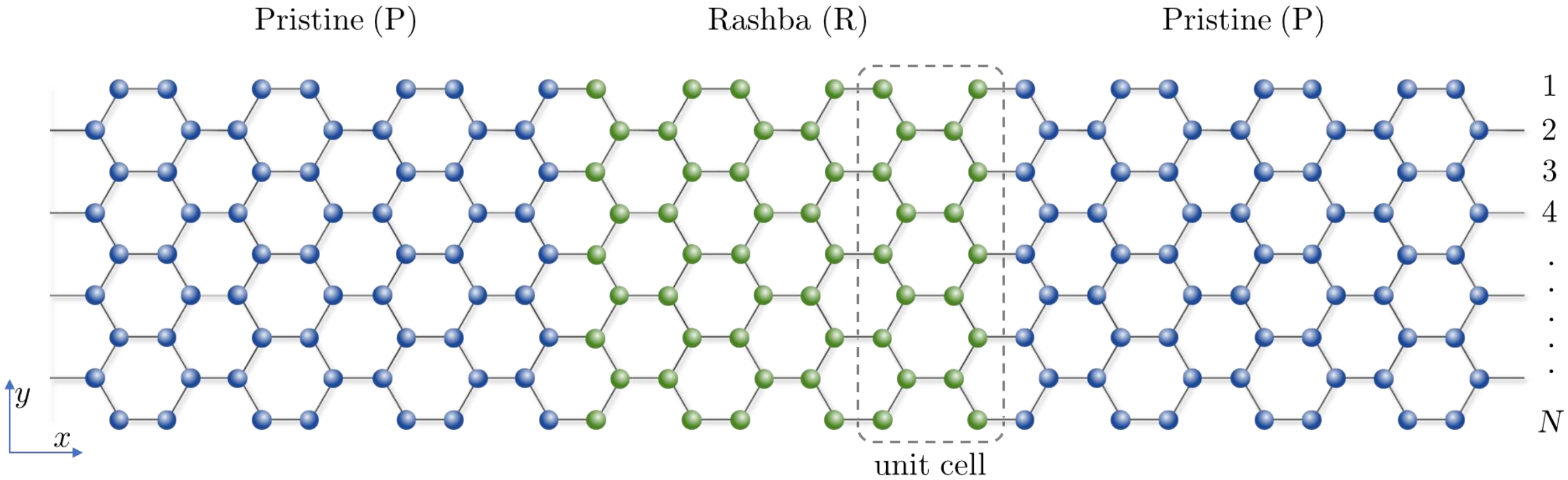

FIG. 1. (Color online) The P-R-P heterostructure. Only the hopping between two green atomic sites contains Rashba SOC.

We consider a heterojunction of three armchair honeycomb nanoribbons of the width $N$ as illustrated in Fig. 1. The Hamiltonian of the P-R-P heterostructure is

$$\hat{H} = \sum_{\substack{\langle i,j\rangle \\ \alpha}} t_0 \hat{c}_{i\alpha}^{\dagger} \hat{c}_{j\alpha} + \sum_{\substack{\langle i,j\rangle \in \mathrm{RGNR} \\ \alpha\beta}} i t_R \hat{c}_{i\alpha}^{\dagger} \boldsymbol{z} \cdot (\boldsymbol{\sigma}_{\alpha\beta} \times \boldsymbol{\delta}_{ij}) c_{j\beta} + \mathrm{h.c.},$$

where $\langle i,j\rangle$ includes all configurations of the nearest-neighbor sites, $\alpha$ and $\beta$ are the spin indices, $c_{i\alpha}^{\dagger}(c_{i\alpha})$ is the creation (annihilation) operator with spin $\alpha$ at site $i$, $t_0$ is the nearest-neighbor hopping strength, $t_R$ is the RSOC strength, $\boldsymbol{\sigma}$ are the Pauli matrices, and $\boldsymbol{\delta}_{ij}$ is the unit vector pointing in the direction from site $i$ to site $j$. Hereafter, we set the nearest-neighbor hopping strength as the unit of energy, $t_0 = 1$. Also, we apply periodic boundary condition on the ends of the pristine armchair nanoribbon to prevent the possible edge states on the rightmost and leftmost zigzag edges.

For $t_R = 0$, the PRP heterostructure reduces to a pristine armchair nanoribbon. The pristine armchair nanoribbon is gapped unless the width $N$ of the armchair nanoribbon satisfy the condition $N = 3M - 1$ with an integer $M > 0$ [34]. Previous studies have shown analytically that the band gap closes every three ribbon widths [35, 36]. For armchair nanoribbon, the edge boundary condition discretizes the wavenumber in the $y$-direction. When the allowed wavenumber hit the Dirac points, which corresponds to the width $N = 3M - 1$, then the nanoribbon becomes metallic [36]. Hence, for $N = 2, 5, 8, 11,$ and so on, the armchair nanoribbon is gapless.

Notably, the additional midgap states emerge within the gap as $t_R$ exceeds a certain number. Fig. 2 shows the energy spectrum and the density of states for the case $N = 6$, for different values of $t_R$. In stark contrast to Fig. 2(a) for $t_R = 0.4$, Fig. 2(c) for $t_R = 0.8$ clearly shows the existence of four midgap states, which are two-fold spin degeneracy at energies $0^+$ and $0^-$. This leads to the pronounced peak at the vicinity of $E = 0$ in the density of state shown in Fig. 2(d).

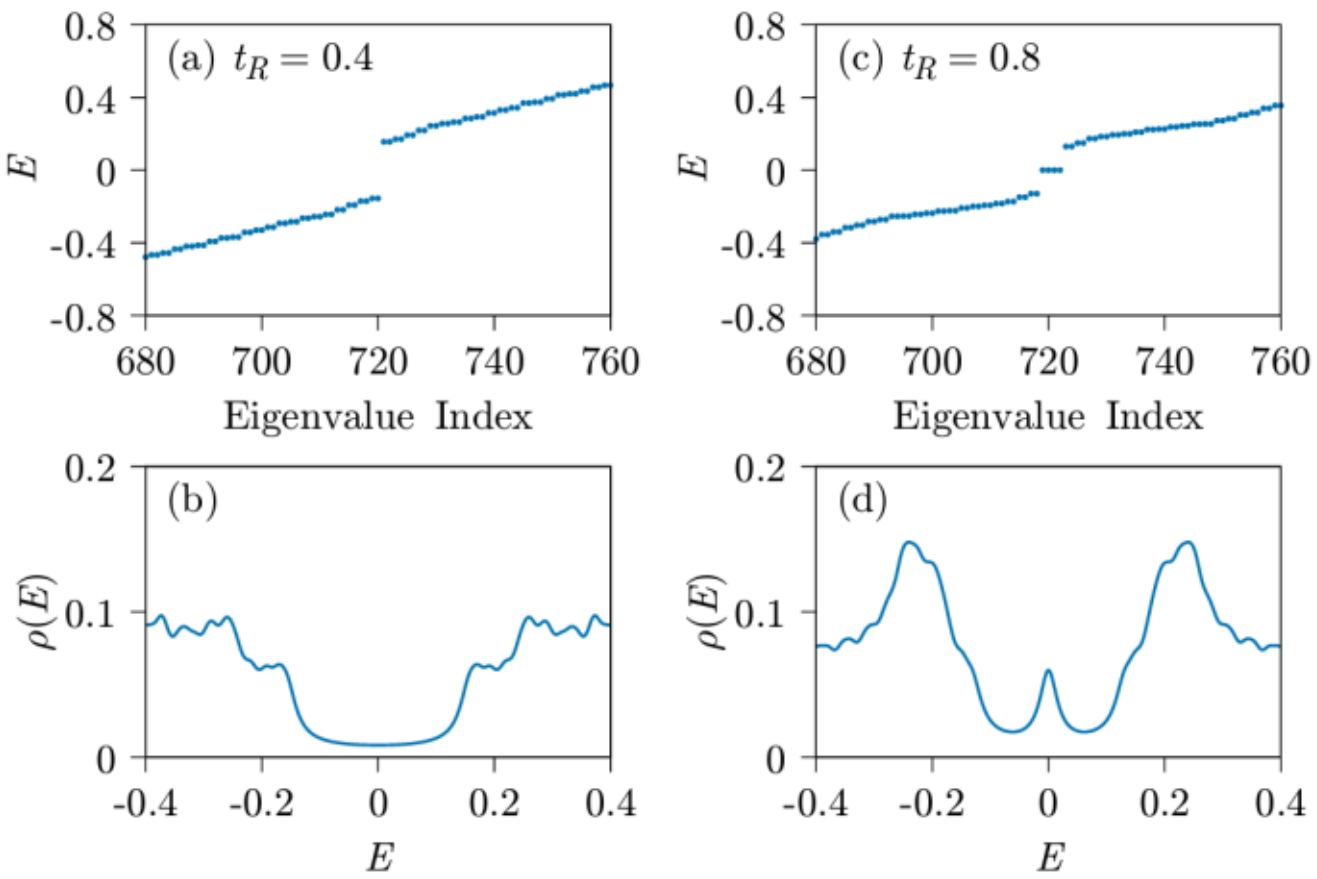

FIG. 2. (a)(c) Energy levels for $N = 6$ heterostructure armchair nanoribbon with $t_R = 0.4$ and $0.8$. The energy gap is $0.26$, $0.31$ respectively. (b)(d) The density of state for $t_R = 0.4$ and $0.8$.

We further explore the local density map for the heterostructure in Fig. 3(a), obtained by integrating the distribution of all mid-gap states. The local density at each point is given by summing the spin-up and spin-down probabilities. We found that these mid-gap states are exponentially localized at the interface between prinstine and Rashba regions. In addition, when the size of the RSOC region is varied, leading to different RSOC–pristine interfaces, such as zigzag (Fig. 3(b)) and bearded (Fig. 3(c)) interfaces, we can still observe the existence of four midgap edge states. This robustness to a variety of interfaces suggests that these states have a topological origin.

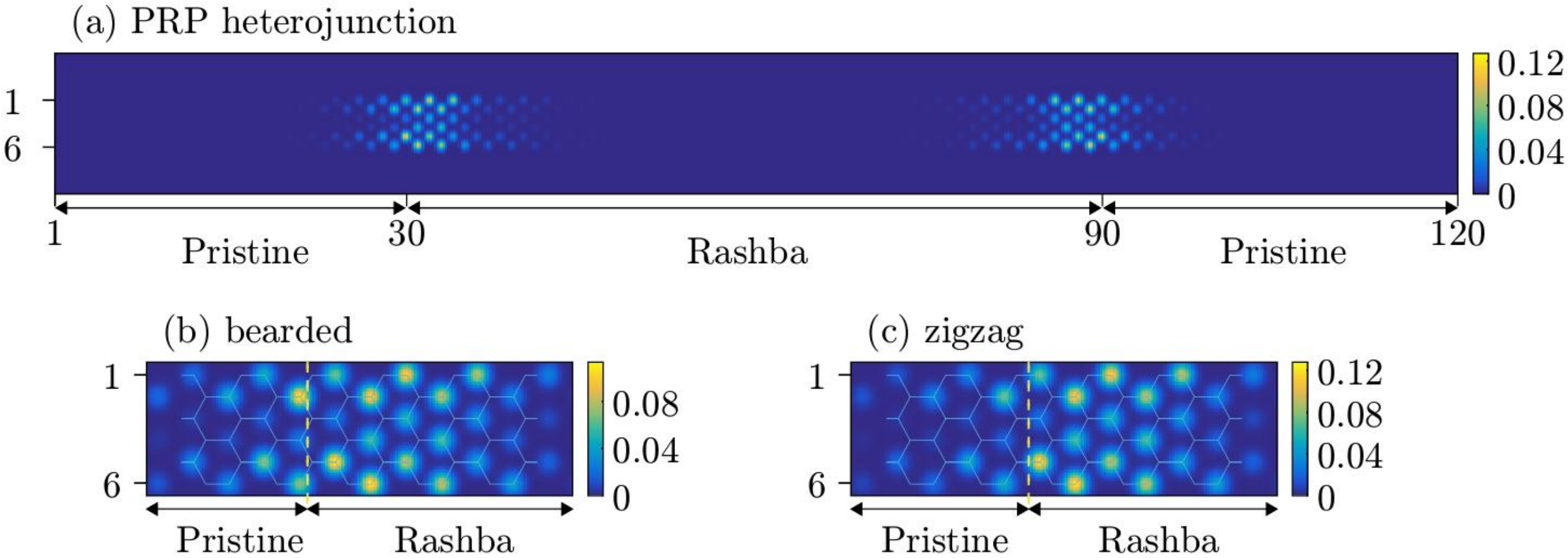

FIG. 3. (Color online) (a). The spatial probability distribution of all midgap states for $N = 6$ and $t_R = 0.8$. (b-c). Distribution plot for different type of interfaces between pristine and Rashba segments.

Motivated by the emergence of robust boundary states, we investigate the topological properties of an armchair nanoribbon uniformly covered by RSOC. We consider an infinite RSOC armchair nanoribbon with width $N$, whose unit cell contains $2N$ atomic sites, as shown in Fig. 1. Due to the translational symmetry along the direction parallel to the edges, the Bloch momentum $k$ is a good quantum number and the Hamiltonian of the quasi-1-D system can be written in a fully block-off-diagonal matrix

$$\mathcal{H} = \begin{pmatrix} 0 & h_k^\dagger \\ h_k & 0 \end{pmatrix}$$

in the basis $\{|A_1^\uparrow\rangle, \cdots, |A_N^\downarrow\rangle, |B_1^\uparrow\rangle, \cdots, |B_N^\downarrow\rangle\}$, where A and B denote the sublattice and $h_k$ is a $2N \times 2N$ matrix. For chiral (sublattice) symmetric system, its 1-D topological properties can be characterized by the winding number [37]

$$W = \int_{-\pi}^{\pi} \frac{dk}{2\pi i} \mathrm{Tr}[h_k^{-1} \partial_k h_k] = \int_{-\pi}^{\pi} \frac{dk}{2\pi i} \partial_k \log[\mathrm{Det}(h_k)].$$

The winding number equals to the number of times the trajectory of $\det(h_k)$ winds around the origin in the complex plane.

For $N = 6$, in the absence of RSOC, the bulk armchair nanoribbon exhibits a finite energy gap (Fig. 4(a)). The trajectory of $\mathrm{Det}(h_k)$ in the complex plane encircles the origin twice, corresponding to a winding number $W = 2$ (Fig. 4(d)). As the RSOC strength $t_R$ increases, the gap gradually decreases. At $t_R = 0.624$, the gap closes (Fig. 4(b)), and the trajectory of $\mathrm{Det}(h_k)$ touches the origin, rendering the winding number ill-defined (Fig. 4(e)). Upon further increasing $t_R$, the gap reopens (Fig. 4(c)), and the trajectory of $\mathrm{Det}(h_k)$ encloses the origin four times, yielding a winding number $W = 4$ (Fig. 4(f)). These results demonstrate that Rashba SOC drives a topological phase transition in armchair nanoribbon, characterized by a change in winding number from $W = 2$ to $W = 4$.

We note that when the $z$-axis values are obtained from the periodic function $\sin(k)$, the winding number can be interpreted as a two-dimensional projection of a three-dimensional unknot trajectory, represented by the red curves in Figs. 4(d)–4(f). For $t_R = 0$, the corresponding three-dimensional curve forms a torus knot $\mathcal{T}(3,1)$, where the trajectory winds three times around the torus hole and once around the tube.

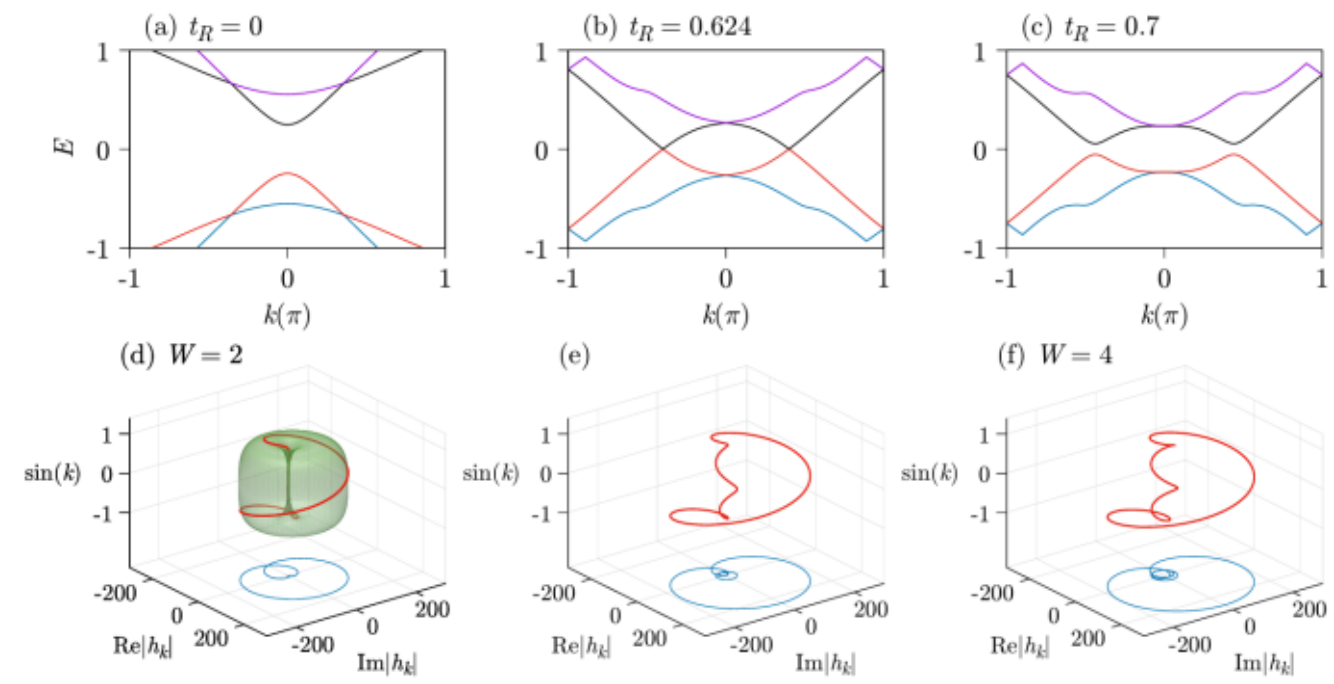

FIG. 4. (Color online) (a) – (c) Four energy bands nearest to the Fermi level with $t_R = 0,\ 0.624$ and $0.7$. In $x$ axis, $k$ are in units of $\pi$. (d) – (f) The path of determinant of $h_k$ in the complex plane as $k$ travels through the whole Brillouin zone with $t_R = 0, 0.624$ and 0.7. The number of times $h_k$ path includes origin point corresponds to the winding number $W$.

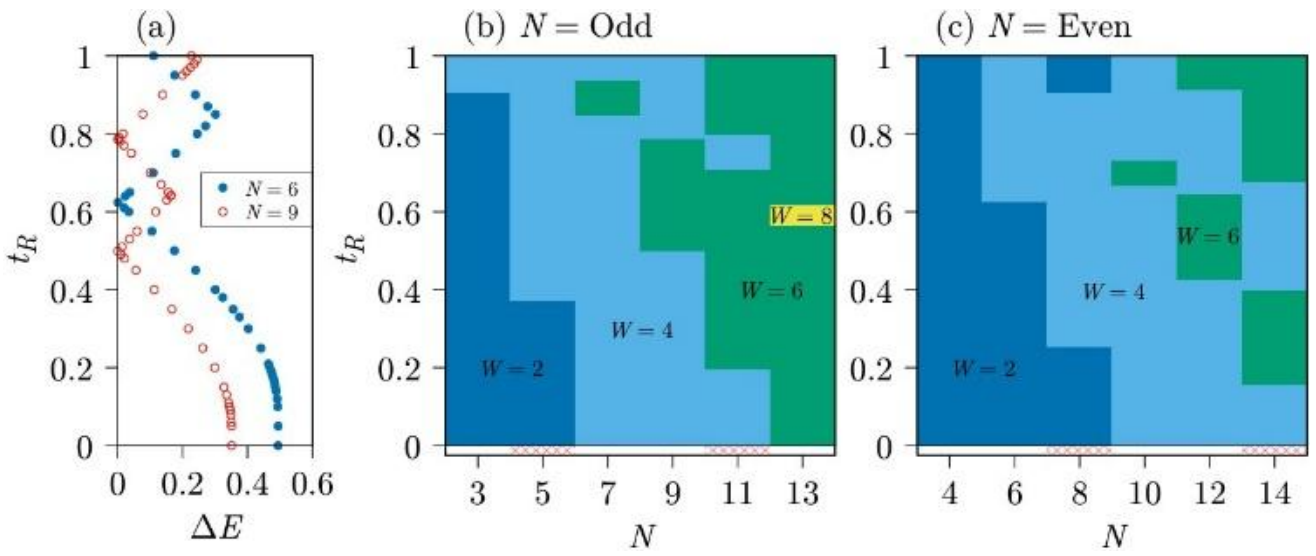

FIG. 5. (Color online) (a) Energy gap relation versus the RSOC strength $t_R$ for $N = 6$ and $N = 9$ nanoribbons. The winding number phase diagram for RSOC strength $t_R$ and the width $N$, for $N$ is (b) odd and (c) even numbers. Regions printed in different colors represent different winding number, and slash area corresponds to a gapless armchair nanoribbon. The phases transition occurred for $N = 3M - 1$, and it's only valid for small $t_R$.

In Fig. 5(a), we show the energy gap versus the RSOC strength $t_R$ for $N \neq 3M - 1$ nanoribbons. When increasing the RSOC strength, the energy gap quadratically decreases at small $t_R$. For $t_R \neq 0$, new Dirac points appear at the vicinity of the original Dirac points with the distances $\sim t_R^2$ for small strength of $t_R$ [38]. As increase $t_R$, the allowed wavenumbers in the $y$-direction moves close to the new Dirac points, the size of gap decreases quadratically. As a consequence, we obtain a quadratically relation of the gap versus the Rashba strength.

Fig. 5(b) and 5(c) show the topological phase diagram as a function of the armchair nanoribbon width $N$ and RSOC strength $t_R$. When $t_R = 0$, our winding number calculations are consistent with the result reported in [8, 9], where armchair graphene nanoribbons (AGNRs) with odd and even widths are classified using different topological categories because the corresponding unit cells exhibit different spatial symmetries [8]. Upon turning on RSOC, the originally gapless nanoribbons with $N = 2, 5, 8, 11, 14$ become gapped and possess a nontrivial winding number. Moreover, when increasing $t_R$, the winding number changes only by two at each transition. For instance, the winding number for $N = 6$ is equal to $2$ when $t_R = 0.4$ and becomes $4$ when $t_R = 0.8$.

In the PRP heterostructure, the pristine and Rashba regions share the same width but are characterized by different winding numbers due to the presence or absence of Rashba spin–orbit coupling. For example, as $N = 6$, the winding number in the pristine region is $W = 2$. By tuning the strength of RSOC $t_R$ on the embedded region, the winding number can change from $W = 2$ to $W = 4$. Thus, the difference of winding number between two regions gives rise to the boundary states localized at the interface (shown in Fig. 2). According to bulk-edge correspondences, this winding number difference leads to the emergence of four symmetry-protected edge states (shown in Figs. 3).

In this work, we have demonstrated that Rashba spin–orbit coupling can serve as an effective and tunable mechanism for generating topological phases in geometrically identical nanoribbon systems. By studying a P–R–P heterostructure, we show that increasing RSOC drives gap closing and reopening associated with changes in the winding number, leading to symmetry-protected interface states consistent with bulk–edge correspondence. The resulting edge modes emerge solely from the contrast in topological invariants between regions with and without Rashba coupling, without requiring modifications of lattice geometry or ribbon width.

This mechanism fundamentally differs from previously explored routes to nanoribbon topology, where edge states arise from junctions between structurally distinct ribbons. Instead, our results establish that spin–orbit interaction alone can induce topological phase transitions within a single geometric platform, providing a dynamically controllable route toward interface topology. Notably, whereas RSOC is commonly understood to suppress topological insulating phases in honeycomb lattice-based systems, we show that in reduced-dimensional geometries it can instead open energy gaps and generate additional topological edge states.

Because the proposed mechanism relies only on bipartite lattice structure and tunable spin–orbit interaction, it is broadly applicable beyond honeycomb nanoribbons and may be realized in electronic, photonic, and synthetic lattice systems. Our findings therefore identify Rashba spin–orbit coupling as a general tool for engineering geometrically preserving yet topologically tunable low-dimensional quantum systems.

## ACKNOWLEDGMENTS

We thank Ion Cosma Fulga for useful discussion. The authors would like to thank "Higher Education Sprout Project" of National Taiwan Normal University and the Ministry of Education (MOE), Taiwan. J.-S.Y. acknowledges support from the National Science and Technology Council (NSTC), Taiwan, under Grant No. NSTC 113-2112-M-003-015, and from TG 3.2 of NCTS at NTU. Y.-R.C thanks the support from National Science and Technology Council (NSTC), grant No. NSTC 114-2635-M-003-001-MY3.